\documentclass[twocolumn,tighten,trackchanges]{aastex63}
\usepackage{amsmath,amssymb,amstext,bm}
\usepackage{graphicx, enumerate}

\submitjournal{ApJ}
\usepackage{float}
\usepackage{xcolor}
\usepackage{soul}

\defcitealias{2011ApJ...740..117E}{Paper I}
\defcitealias{2014ApJ...790..130B}{Paper II}
\defcitealias{1995ApJ...438..763G}{GS95}
\defcitealias{2016MNRAS.461.1227K}{KLP16}
\defcitealias{2017MNRAS.464.3617K}{KLP17}

\shorttitle{Anisotropy and MHD modes in the ISM}
\shortauthors{Hern\'andez-Padilla et al.}

\begin{document}

\title{Velocity Centroids Anisotropy and the Signature of
 different MHD Modes in the Turbulent ISM}

\correspondingauthor{D. Hern\'andez-Padilla}
\email{david.hernandez@correo.nucleares.unam.mx}

\author[0000-0001-9574-1319]{D. Hern\'andez-Padilla}
\affiliation{
Instituto de Ciencias Nucleares, Universidad Nacional Aut\'onoma de M\'exico,
Apartado Postal 70-543, 04510 Ciudad de M\'{e}xico, M\'{e}xico}

\author[0000-0001-7222-1492]{A. Esquivel}
\affiliation{
Instituto de Ciencias Nucleares, Universidad Nacional Aut\'onoma de M\'exico,
Apartado Postal 70-543, 04510 Ciudad de M\'{e}xico, M\'{e}xico}
\affiliation{Instituto de Astronom\'ia Te\'orica y Experimental, CONICET - UNC, Laprida 854, X5000BGR C\'ordoba, Argentina}

\author[0000-0002-7336-6674]{A. Lazarian}
\affiliation{
Astronomy Department, University of Wisconsin-Madison,
475 N. Charter Street, Madison, WI 53706, USA}
\affiliation{Center for Computation Astrophysics, Flatiron Institute, 162 5th Ave, New York, NY 10010}

\author[0000-0002-7998-6823]{D. Pogosyan}
\affiliation{Physics Department, University of Alberta, Edmonton, Alberta T6G 2E1, Canada}

\author[0000-0002-5402-3107]{D. Kandel}
\affiliation{Stanford University, Stanford, CA 94309, USA}

\author[0000-0003-1725-4376]{J. Cho}
\affiliation{
Department of Astronomy and Space Science, Chungnam
National University, Daejeon 305-764, Korea
}



\begin{abstract}

Magnetic turbulence is anisotropic as the   directions of motion are constrained by the magnetic field. Such anisotropy can be observed in velocity centroids obtained from spectroscopic observations.
We use magnetohydrodynamics (MHD) simulations to produce synthetic spectroscopic observations (position-position-velocity data) and study the anisotropy in the structure function of velocity centroid maps. We decomposed the velocity in the simulations into Alfv\'en, slow and fast-modes and studied how each of them contribute to the observed anisotropy. We found that when the angle between the line of sight and the mean magnetic field is large the Alfv\'en-mode dominates the observed anisotropy, while for smaller angles the anisotropy is not large enough to be used to probe the magnetization of the media, and it is dominated by the slow-mode. Our results are in fair agreement with the theoretical predictions in \citet{2016MNRAS.461.1227K,2017MNRAS.464.3617K}.

\end{abstract}

\keywords{
ISM: general --- ISM: structure --- magnetohydrodynamics (MHD) ---
radio lines: ISM  --- turbulence}


\section{Introduction}\label{sec:intro}

Given the large Reynolds numbers (defined as the ratio of inertial to viscous forces) prevailing in the interstellar medium (ISM), its natural state is turbulent. Evidence of this turbulence has been observed in scales that range from kiloparsecs to sub-astronomical units \citep{1995ApJ...443..209A, 2010ApJ...710..853C, 2016ApJ...824..113X, 2016ApJ...832..199X}. The implications of such turbulence are just as ubiquitous in the ISM. For instance, it is of paramount importance in the process of star formation \citep{2004RvMP...76..125M, 2007prpl.conf...63B,2007ARA&A..45..565M},
in the acceleration and propagation of cosmic-rays \citep{2004ApJ...614..757Y}, heat transfer \citep{2001ApJ...562L.129N, 2006ApJ...645L..25L}, and in many other transport phenomena in the ISM \citep{2004ARA&A..42..211E}.

At the same time, there is a magnetic field that permeates the ISM and it has been recognized as well as an essential ingredient in many processes that take place in the ISM. Turbulence and magnetic fields are intertwined, and their interplay is complex \cite[see][]{2013SSRv..178..163B,2019tuma.book.....B}.

The MHD turbulence has been studied for decades \citep[see][]{2003matu.book.B}. However, the radical shift of the understanding of magnetohydrodynamic (MHD) turbulence can be traced to the model suggested  in  \citep[][hereafter \citetalias{1995ApJ...438..763G}]{1995ApJ...438..763G}.  That model predicted that while motions perpendicular to the magnetic field follow a \citet{1941DoSSR..30..301K} type cascade, the motions along the magnetic field are different.  The coupling between motions perpendicular and parallel to the magnetic field in \citetalias{1995ApJ...438..763G} is achieved by the so-called {\it critical balance} condition which in the original formulation was $k_\perp v_\perp \sim k_\parallel v_\mathrm{A}$.  This formulation assumed that the anisotropy is in terms of mean field. In fact, the further work clarified that this is not true. Indeed, adopting the picture of turbulent reconnection in \citep{1999ApJ...517..700L} it is possible to show that the reconnection is fast enough to enable eddy-like motions perpendicular to the direction of magnetic field {\it local} to the eddies. As a result instead of $k_{\bot}$ and $k_\|$ measured in respect to the global mean field, one should use the $l_{\bot}^{-1}$ and $l_\|^{-1}$ where the latter two quantities are measure in respect to the local direction of magnetic field. In other words, the MHD turbulence should be described not in $k$-space, but in real space. In terms of reconnecting eddies, the Kolmogorov statistics of perpendicular motions follows trivially, as the corresponding eddies are not constrained by magnetic field tension. Combining this with the critical balance condition written in real space one easilty gets $l_\parallel \sim l_\perp^{2/3}$. This  {\it scale dependent} anisotropy with respect with the {\it local} magnetic field was later confirmed by numerical simulations \citep{2000ApJ...539..273C,2001ApJ...554.1175M,2002ApJ...564..291C}.

The \citetalias{1995ApJ...438..763G} is the theory of incompressible MHD turbulence. In a realistic compressible case, MHD turbulence can be modeled as a superposition of three fundamental modes, the Alfv\'en, slow and fast modes \citep[see][]{2001ApJ...562..279L}. This was numerically demonstrated in \citep{2002PhRvL..88x5001C, 2003MNRAS.345..325C,2010ApJ...720..742K}, where these the properties of these modes were explored and quantified. It was found there that the transfer of energy between the modes is relatively small, which allows one to talk about the three cascades of modes.  The compressible nature of ISM turbulence is contained in the slow and fast modes. The interplay between the compressible and incompressible (Alfv\'en) modes has far-reaching consequences. For instance, turbulence can induce star formation by forming overdense regions via compressible modes, but at the same time the incompressible part of turbulence can produce additional support against gravity collapse. Compressible turbulence also plays a crucial role in cosmic ray diffusion and acceleration \citep{2004ApJ...614..757Y,2011ApJ...728...60B, 2016ApJ...826..166X}.

Several methods have been also developed to study ISM turbulence from observations, using for instance the line-widths in spectroscopic observations \citep{1981MNRAS.194..809L,1992MNRAS.256..641L,1984ApJ...277..556S,1987ASSL..134..349S}, velocity centroids \citep{1951ZA.....30...17V, 1958RvMP...30.1035M,1985ApJ...295..466K,1985ApJ...295..479D,1994ApJ...429..645M}, or measuring the fluctuations of electron density from scintillations \citep{1989MNRAS.238..963N,1990ApJ...353L..29S}.
In the last couple of decades several new techniques have been put forward to obtain turbulence information from the widely available radio spectroscopic observations. Among these techniques the Velocity Channel Analysis and Velocity Coordinate Spectrum  \citep[VCA and VCS respectively][]{2000ApJ...537..720L,2004ApJ...616..943L,2002ASPC..276..182L,2003MNRAS.342..325E,2005ApJ...631..320E,2009ApJ...693.1074C} are the technique based on the analytical theory of the mapping of velocity fluctuations from the real space into the Position-Position-Velocity (PPV) space. The Spectral Correlation Function \citep[SCF;][]{1999ApJ...524..887R,2001ApJ...547..862P} is in many respects an empirical analog of the VCA, the Principal component Analysis \citep[PCA;][]{1997ApJ...475..173H,2002ApJ...566..289B} is another empirical technique for turbulence studies.

Magnetic fields have been traditionally studied with a number of observational techniques suited particularly for that purpose \citep[see reviews by][]{1967ARA&A...5..167V,1976ARA&A..14....1H,2012ARA&A..50...29C}. Among them the Zeeman splitting is intended for studies of the parallel to the line of sight component of magnetic field, while dust polarization reveals the plane of the sky direction of magnetic field.

Due to better understanding of the nature of MHD turbulence it became possible to propose statistical techniques that can also trace the plane of sky direction of magnetic field. This, for instance, can be obtained measuring the anisotropy in the structure functions of centroids of velocity \citep{2002ASPC..276..182L, 2003MNRAS.342..325E, 2005ApJ...631..320E}. It is also important that besides the direction of the mean magnetic field in the plane of the sky, one can also obtain information about the degree of media magnetization, i.e. about the Alfv\'en Mach number $M_A=V_L/V_A$, where $V_L$ is a turbulent injection velocity and $V_A$ is the Alfv\'en velocity, from the degree of anisotropy of velocity centroids (\citealt{2011ApJ...740..117E,2014ApJ...790..130B}; hereafter \citetalias{2011ApJ...740..117E}, and \citetalias{2014ApJ...790..130B}, respectively), or, equivalently, of velocity channel maps \citep{2015ApJ...814...77E}.

The theory of emissivity fluctuations in PPV space proposed by \citet{2000ApJ...537..720L,2004ApJ...616..943L} has been recently extended to account for the anisotropy in velocity channels (\citealt*{2016MNRAS.461.1227K}; hereafter \citetalias{2016MNRAS.461.1227K}) and in velocity centroids (\citealt*{2017MNRAS.464.3617K}; hereafter \citetalias{2017MNRAS.464.3617K}), arising from the different fast, slow, and Alfv\'en MHD modes.
\citetalias{2016MNRAS.461.1227K} and \citetalias{2017MNRAS.464.3617K} papers are important because the provide with a theoretical framework for studying the anisotropy using velocity centroids. Moreover, they allow to determine the contribution of different MHD modes. This provides an important insight in the picture of the magnetized turbulence in the ISM. With a better understanding of the behavior of the different modes one can improve the current techniques, or develop new ones\footnote{For instance, on the basis of the understanding of the nature of MHD turbulence and the relation between the fluctuations in the PPV and real space the Velocity Gradients Technique (VGT) \citep[VGT;][]{2017ApJ...835...41G,2017ApJ...837L..24Y,2018ApJ...865...54Y,2019MNRAS.483.1287G} has been proposed.} with better accuracy by removing some of the modes \citep{2018ApJ...865...54Y}.
In this work we revisit our previous studies of the anisotropy in velocity centroids done in  \citetalias{2011ApJ...740..117E} and \citetalias{2014ApJ...790..130B}, but we analyze each of the MHD modes separately and compare the results with the theoretical predictions in \citetalias{2017MNRAS.464.3617K}.

The paper is organized as follows: we describe the set of numerical models used in Section \ref{sec:models}, we review the anisotropy of velocity centroids and layout our method in Section \ref{sec:centroids}. The results are presented in Section \ref{sec:results}, followed by a summary in Section \ref{sec:sum}.
\section{Models}\label{sec:models}

We use a grid of MHD simulations of compressible,  isothermal,  fully developed turbulence. The grid is similar to the one used in \citet{2015ApJ...814...77E}, but all the models have been updated to the same resolution (of $512^3$ cells). The simulations are obtained with a second-order hybrid essentially non-oscillatory (ENO) method that solves the ideal MHD equations with a turbulence forcing term, in a periodic Cartesian box \citep[see ][]{2002PhRvL..88x5001C}. The driving  is purely solenoidal and it is imposed in Fourier space at a fixed wave-number $k=2.5$ (which corresponds to a scale that is $1/2.5$ of the computational domain). The magnetic field is composed by a uniform background $\bm{B}_0$ plus a fluctuating part $\bm{b}$. The simulations start with a homogeneous medium of constant density $\rho_0=1$, and a uniform magnetic field aligned with the $x$-axis ($\bm{B}_0={B}_0\, \bm{\hat{x}}$, and initially $\bm{b}=0$).

The simulations are evolved, constantly driven, until they reach a stationary state, where the rms velocity is of order unity (see Table \ref{tab:models}). Each model can be characterized by two parameters: the sonic Mach number, and the Alfv\'enic Mach numbers, $\mathcal{M}_\mathrm{s}=\left\langle V_\mathrm{L}/c_\mathrm{s}\right\rangle$, and $\mathcal{M}_\mathrm{A}=\left\langle V_\mathrm{L}/v_\mathrm{A}\right\rangle$, respectively; where $V_\mathrm{L}=v_\mathrm{rms}$ is the velocity at the injection scale, $c_\mathrm{s}=\sqrt{P/\rho}$ the sound speed, $v_\mathrm{A}=\vert\bm{B}\vert/\sqrt{4\pi \rho}$ the Alfv\'en speed, and $\left\langle\dots\right\rangle$ denotes an average over the entire domain. These parameters are in turn controlled by the values of the initial Alfv\'en speed $v_\mathrm{A,0}=\vert\bm{B}_0\vert/\sqrt{4\pi \rho_0}$, and the initial gas pressure $P_\mathrm{gas,0}$.

When the simulations reach a stationary state, the fluctuations of the magnetic field can be of the order of the uniform field, but the mean field remains aligned in the original orientation (along $x$). The different models are summarized in Table \ref{tab:models}, where the initial conditions and the resulting Mach numbers are listed. The parameters explored cover a wide range of sub-sonic and supersonic, along with sub-Alfv\'enic and super-Alfv\'enic turbulence regimes. In the Table we also include the plasma $\beta=P_\mathrm{gas}/P_\mathrm{mag}$, which also covers the regimes in which the magnetic fields are dynamically dominant ($\beta \ll 1$), or dynamically unimportant ($\beta \gg 1$).

\begin{deluxetable*}{lCCCCCCCCC}
\tablecaption{Grid of MHD simulations.\label{tab:models}}
\tablewidth{0pt}
\tablehead{
    \colhead{Model}
  & \colhead{$v_\mathrm{A,0}$}
  & \colhead{$P_\mathrm{gas,0}$}
  & \colhead{$v_{rms}$}
  & \colhead{$\mathcal{M}_\mathrm{A}$}
  & \colhead{$\mathcal{M}_\mathrm{s}$}
  & \colhead{$\beta$}
  & \colhead{$\left\langle \rho\,u_\mathrm{A}^2\right\rangle$}
  & \colhead{$\left\langle \rho\,u_\mathrm{s}^2\right\rangle$}
  & \colhead{$\left\langle \rho\,u_\mathrm{f}^2\right\rangle$} \\
  & & & & & & & \colhead{(\%)} & \colhead{(\%)} & \colhead{(\%)}
}
\startdata
M1  &  0.1  &  0.01  & \sim 0.79 & \sim 7.90  & \sim 7.90  & 2      & \sim 43.4 & \sim 41.1 & \sim 15.5 \\
M2  &  0.1  &  0.10  & \sim 0.78 & \sim 7.80  & \sim 2.47  & 20     & \sim 42.7 & \sim 46.9 & \sim 10.4 \\
M3  &  0.1  &  1.00  & \sim 0.77 & \sim 7.71  & \sim 0.77  & 200    & \sim 48.0 & \sim 48.4 & \sim  3.6 \\
M4  &  0.1  &  2.00  & \sim 0.75 & \sim 7.49  & \sim 0.53  & 400    & \sim 48.8 & \sim 47.9 & \sim  3.3 \\
\hline
M5  &  0.5  &  0.01  & \sim 0.72 & \sim 1.43  & \sim 7.16  & 0.08   & \sim 40.8 & \sim 44.5 & \sim 14.7 \\
M6  &  0.5  &  0.10  & \sim 0.69 & \sim 1.37  & \sim 2.17  & 0.8    & \sim 42.8 & \sim 44.9 & \sim 12.3 \\
M7 	&  0.5  &  1.00  & \sim 0.67 & \sim 1.34  & \sim 0.67  & 8      & \sim 42.8 & \sim 53.8 & \sim  3.4 \\
M8	&  0.5  &  2.00  & \sim 0.66 & \sim 1.32  & \sim 0.47  & 16     & \sim 42.1 & \sim 55.5 & \sim  2.5 \\
\hline
M9  &  1.0  &  0.01  & \sim 0.75 & \sim 0.75  & \sim 7.54  & 0.02   & \sim 45.5 & \sim 44.5 & \sim 10.0 \\
M10 &  1.0  &  0.10  & \sim 0.72 & \sim 0.72  & \sim 2.28  & 0.2    & \sim 47.5 & \sim 43.6 & \sim  8.9 \\
M11 &  1.0  &  1.00  & \sim 0.76 & \sim 0.76  & \sim 0.76  & 2      & \sim 39.5 & \sim 55.2 & \sim  5.3 \\
M12 &  1.0  &  2.00  & \sim 0.77 & \sim 0.77  & \sim 0.54  & 4      & \sim 38.1 & \sim 58.4 & \sim  3.4 \\
\hline
M13 &  2.0  &  0.01  & \sim 0.76 & \sim0.38   & \sim 7.62  & 0.005  & \sim 52.2 & \sim 45.9 &  \sim 1.9 \\
M14 &  2.0  &  0.10  & \sim 0.77 & \sim0.39   & \sim 2.45  & 0.05   & \sim 58.8 & \sim 38.7 &  \sim 2.5 \\
M15 &  2.0  &  1.00  & \sim 0.84 & \sim0.42   & \sim 0.84  & 0.5    & \sim 62.6 & \sim 33.9 &  \sim 3.4 \\
M16 &  2.0  &  2.00  & \sim 0.83 & \sim0.42   & \sim 0.59  & 1      & \sim 61.3 & \sim 34.7 &  \sim 4.0 \\
\hline
M17 &  3.0  &  0.01  & \sim 0.80 & \sim 0.27  & \sim 8.05  & 0.0022 & \sim 63.6 & \sim 35.5 &  \sim 0.9 \\
M18 &  3.0  &  0.10  & \sim 0.81 & \sim 0.27  & \sim 2.57  & 0.022  & \sim 64.0 & \sim 34.6 &  \sim 1.4 \\
M19 &  3.0  &  1.00  & \sim 0.84 & \sim 0.28  & \sim 0.84  & 0.22   & \sim 64.9 & \sim 30.5 &  \sim 4.6 \\
M20 &  3.0  &  2.00  & \sim 0.82 & \sim 0.27  & \sim 0.58  & 0.44   & \sim 62.1 & \sim 30.7 &  \sim 7.2 \\
\hline
M21 &  5.0  &  0.01  & \sim 0.86 & \sim 0.17  & \sim 8.61  & 0.0008 & \sim 67.5 & \sim 31.6 & \sim  0.9 \\
M22 &  5.0  &  0.10  & \sim 0.85 & \sim 0.17  & \sim 2.70  & 0.008  & \sim 68.5 & \sim 30.6 & \sim  0.9 \\
M23 &  5.0  &  1.00  & \sim 0.83 & \sim 0.17  & \sim 0.83  & 0.08   & \sim 58.5 & \sim 28.8 & \sim 13.5 \\
M24 &  5.0  &  2.00  & \sim 0.81 & \sim 0.16  & \sim 0.57  & 0.16   & \sim 47.7 & \sim 20.2 & \sim 32.1
\enddata
\tablecomments{ $\mathcal{M}_\mathrm{A}= v_\mathrm{rms}/v_\mathrm{A}$, $\mathcal{M}_\mathrm{s}= v_\mathrm{rms}/c_\mathrm{s}$, and $\beta = P_\mathrm{gas}/P_\mathrm{mag} =2 \left( \mathcal{M}_\mathrm{A}^2/ \mathcal{M}_\mathrm{s}^2\right) $. All the models have a resolution of $512^3$ cells.}
\end{deluxetable*}

\subsection{MHD mode decomposition}\label{sec:modes}

In a compressible and magnetized plasma there are three waves (or {\it modes}) that propagate, namely the  Alfv\'en, slow and fast MHD waves. Since each of them has a different anisotropy, we follow the procedure in \citet{2002PhRvL..88x5001C,2003MNRAS.345..325C} to decompose the original velocity field into each of the MHD modes. The decomposition is obtained by projecting the Fourier components of the velocity onto the direction of the displacement vectors of the Alfv\'en, slow, and fast modes $\bm{\hat{\xi}}_\mathrm{A}$, $\bm{\hat{\xi}}_\mathrm{s}$, $\bm{\hat{\xi}}_\mathrm{f}$, respectively.
The directions of the plasma displacement are defined as:
\begin{equation}
  \label{eq:disA}
  \bm{\hat{\xi}}_\mathrm{A} =  -
   \bm{\hat{\varphi}}=\bm{\hat{k}}_\perp \times \bm{\hat{k}}_\parallel,
\end{equation}
\begin{equation}
  \label{eq:disS}
  \bm{\hat{\xi}}_\mathrm{s} \propto
   \left(-1+\alpha-\sqrt{D}\right)\,k_\parallel\,\bm{\hat{k}}_\parallel +
   \left( 1+\alpha-\sqrt{D}\right)\,k_\perp\,    \bm{\hat{k}}_\perp,
\end{equation}
and
\begin{equation}
  \label{eq:disF}
  \bm{\hat{\xi}}_\mathrm{f} \propto
   \left(-1+\alpha+\sqrt{D}\right)\,k_\parallel\,\bm{\hat{k}}_\parallel +
   \left( 1+\alpha+  \sqrt{D}\right)\,k_\perp\,    \bm{\hat{k}}_\perp.
\end{equation}
Where $\bm{\hat{k}}_\parallel$, and $\bm{\hat{k}}_\perp$ are the components of $\bm{k}$ (the direction of the velocity in Fourier space), which are parallel and perpendicular to $\bm{B}_\mathrm{ext}=\bm{B}_0$, respectively; $\alpha=c_\mathrm{s}^2/v_\mathrm{A}^2$; $D=(1-\alpha)^2-4\,\alpha\,\cos^2\theta$; $\theta$ being the angle between $\bm{k}$, and $\bm{B}_\mathrm{ext}$ and $\bm{\hat{\varphi}}$ the azimuthal coordinate in Fourier space.\footnote{For further detail see figure 1, and appendix A in \citealt{2003MNRAS.345..325C}} The Fourier projection of each mode is transformed back to real space to obtain the corresponding velocity field. The result is that from each velocity field $\bm{u}$ we have three velocity fields (each one with $x$, $y$, and $z$ components), $\bm{u}_\mathrm{A}$, $\bm{u}_\mathrm{f}$ and $\bm{u}_\mathrm{s}$ (for the Alfv\'en, slow and fast modes, respectively).

We show in the last three columns of Table \ref{tab:models} the percentage of kinetic energy contained in each mode with respect to the original kinetic energy, the values quoted are with respect to the total kinetic energy $\left\langle\rho\,u^2\right\rangle/2$.
We can see that on most of the kinetic energy is contained in the Alfv\'en modes, with 52\% of the energy on average considering all the models.
This average becomes $56.4\%$ if we consider only the sub-Alf\'enic models, and close to $43.9\%$ for the super-Alfv\'enic cases.
The slow modes contain globally $40.8\%$ of the kinetic energy, being stronger for  super-Alfv\'enic turbulence ($47.9\%$) and weaker in sub-Alev\'enic models ($37.3\%$). The Fast mode is the weakest of the three, with only $6.9\%$ of the energy in general. We can see an increase on the strength of the fast modes with the Mach number for Models with $v_\mathrm{A,0} \leq 1$.
All this trends are consistent with the findings by \citet{2010ApJ...720..742K}, which used an extension of the method of mode decomposition used here that makes use of wavelets to obtain a local measure of the $\alpha$ and $D$ coefficients in the decomposition.
For $v_\mathrm{A,0} \geq 2$ we see an opposite trend in the the fast modes, in which their strength becomes weaker for higher $\mathcal{M}_\mathrm{s}$, this deserves future investigation.

The percentages described above were obtained using the entire 3D velocity field, irrespective of the direction of the magnetic field.
However, it is important to note that the magnitude of the velocity in each mode does vary significantly with respect of the direction of the magnetic field (hence the anisotropy). In our simulations the $x$-component of the velocity field is parallel to the mean magnetic field, whereas the $y$ and $z$-components are perpendicular to $\bm{B}_0$. This would ultimately be reflected on the observed centroids which only map the velocity projected along a given line of sight (LOS).


\section{Velocity centroids and their anisotropy}
\label{sec:centroids}

As mentioned before, the (scale dependent) anisotropic MHD turbulence described in the model of \citetalias{1995ApJ...438..763G} should only be interpreted within the local magnetic field for each turbulent eddy \citep{1999ApJ...517..700L}.
In observations, however, the resulting local anisotropy can not be studied directly because we do not have access to the three dimensional distribution of the emitting material. For instance, in  spectroscopic observations one obtains instead the distribution of emitting material in velocity at a certain position (or a set of positions) on the plane of the sky. And, since at a given velocity  one can have contribution from any position along the LOS, the anisotropy measured would correspond to a {\it global} magnetic field. In \citetalias{2011ApJ...740..117E} and \citetalias{2014ApJ...790..130B} we study the anisotropy in velocity centroids in such a global frame of reference, and found the anisotropy to be mostly {\it scale independent} within the  inertial  range.

Recently, \citetalias{2016MNRAS.461.1227K} extended the theoretical framework of VCA to study the statistics of velocity centroids from an analytical perspective. The theoretical description  therein, allowed for a natural  follow up to address their anisotropy from an analytical perspective as well in \citetalias{2017MNRAS.464.3617K}.

In view of this progress, in the present work we revisit our previous studies (\citetalias{2011ApJ...740..117E,2014ApJ...790..130B})  of the anisotropy in synthetic observations to compare with the predictions of \citetalias{2017MNRAS.464.3617K}.
We are also interested in exploring the contribution to the observed anisotropy by each of the MHD modes. To do this, we take the results of the MHD simulations and obtain synthetic observations in the form of position-position-velocity (PPV) data cubes.
The PPV data cubes can be treated in the same manner as real observations, for instance we can get 2D maps of integrated intensity, and velocity centroids.
In order to obtain mock observations with different orientations we obtain several synthetic PPV data cubes, rotating the data from the simulations by an angle $\gamma$ with respect to the $y$-axis, and taking the LOS in the direction of the original $x$-axis. Thus, $\gamma$ corresponds to the angle between the LOS and the mean magnetic field.

From the PPV cubes, the integrated intensity (proportional to the column density in the case on an optically thin media with emissivity proportional to density, e.g. cold \ion{H}{1}) can be obtained as
\begin{equation}
    \label{eq:Int}
    I_\mathrm{\gamma}\left(\bm{X} \right)=\int \rho_\mathrm{s,\gamma}\left(\bm{X},v_\mathrm{los} \right)\,dv_\mathrm{los},
\end{equation}
where $\rho_\mathrm{s,\gamma}$ is the density of emitters arranged in PPV space,\footnote{One important technical detail is that we take advantage of the fact that the simulations have periodic boundary conditions to produce rotated PPV cubes that are uniformly sampled (with the same number of points) at each position in the plane of the sky.} obtained for an angle $\gamma$, while $\bm{X}$ is the position in the plane of the sky, and $v_\mathrm{los}$ the velocity projection in the LOS.

As with real observations, from the PPV data one can obtain the velocity centroids as the first moment of the spectral lines:
\begin{equation}
    \label{eq:Cent}
        C_\mathrm{\gamma}\left(\bm{X} \right)=
        \frac{\int \rho_\mathrm{s,\gamma}\left(\bm{X},v_\mathrm{los} \right)\,v_\mathrm{los}\,dv_\mathrm{los}}{I_\mathrm{\gamma}\left(\bm{X} \right)}.
\end{equation}
This is the usual definition of velocity centroids, which we have often termed as {\it normalized} centroids, in order to distinguish them from the {\it unnormalized} centroids which are not divided by the column density \citep{2005ApJ...631..320E}. The normalization decreases the impact of density fluctuations to the observed maps, but it is somewhat difficult to include in an analytical study of the statistics. Thus, the description by  \citetalias{2016MNRAS.461.1227K,2017MNRAS.464.3617K} was cast in terms of unnormalized centroids. Moreover, \citetalias{2016MNRAS.461.1227K,2017MNRAS.464.3617K}, neglected the effects of density fluctuations while constructing the centroids, this is equivalent to compute the centroids with a constant density in Equation (\ref{eq:Cent}), which is also equivalent to consider the mean LOS velocity\footnote{The mean LOS velocity can not be obtained from observations, velocity centroids (a density weighted average) are used as a proxy for it.}:
\begin{equation}
    \label{eq:meanV}
    V_\mathrm{\gamma}\left(\bm{X}\right)=\frac{1}{N_\mathrm{los}}\int \bm{u}\left(\bm{x}\right)\cdot\bm{\hat e}_{los}\,d \ell_\mathrm{los},
\end{equation}
where $N_\mathrm{los}$ is the number of points (computational cells) crossed in each LOS, $\bm{\hat{e}}$ the direction, and $\ell_\mathrm{los}$ denotes the distance also in the direction of the LOS. To gauge the impact of density fluctuations in our work, we also construct maps of mean LOS velocity from the simulations, producing PPV data with constant density (see also \citetalias{2011ApJ...740..117E}).

In order to study the individual contributions to the general anisotropy we use the velocity fields that correspond to the Alfv\'en,  slow, and fast MHD modes. With each of these velocity fields we construct PPV data, and with them maps of mean LOS velocity, and velocity centroids. We will denote these with an additional subscript ``A", ``s", or ``f", for the Alfv\'en, slow, and fast modes, respectively (e.g. $V_\mathrm{A,\gamma}$ is the mean LOS velocity for the Alfv\'en mode integrated in a LOS at an angle $\gamma$ with the direction of the mean magnetic field).

\begin{figure*}[!htp]
  \centering
  \includegraphics[width=\textwidth]{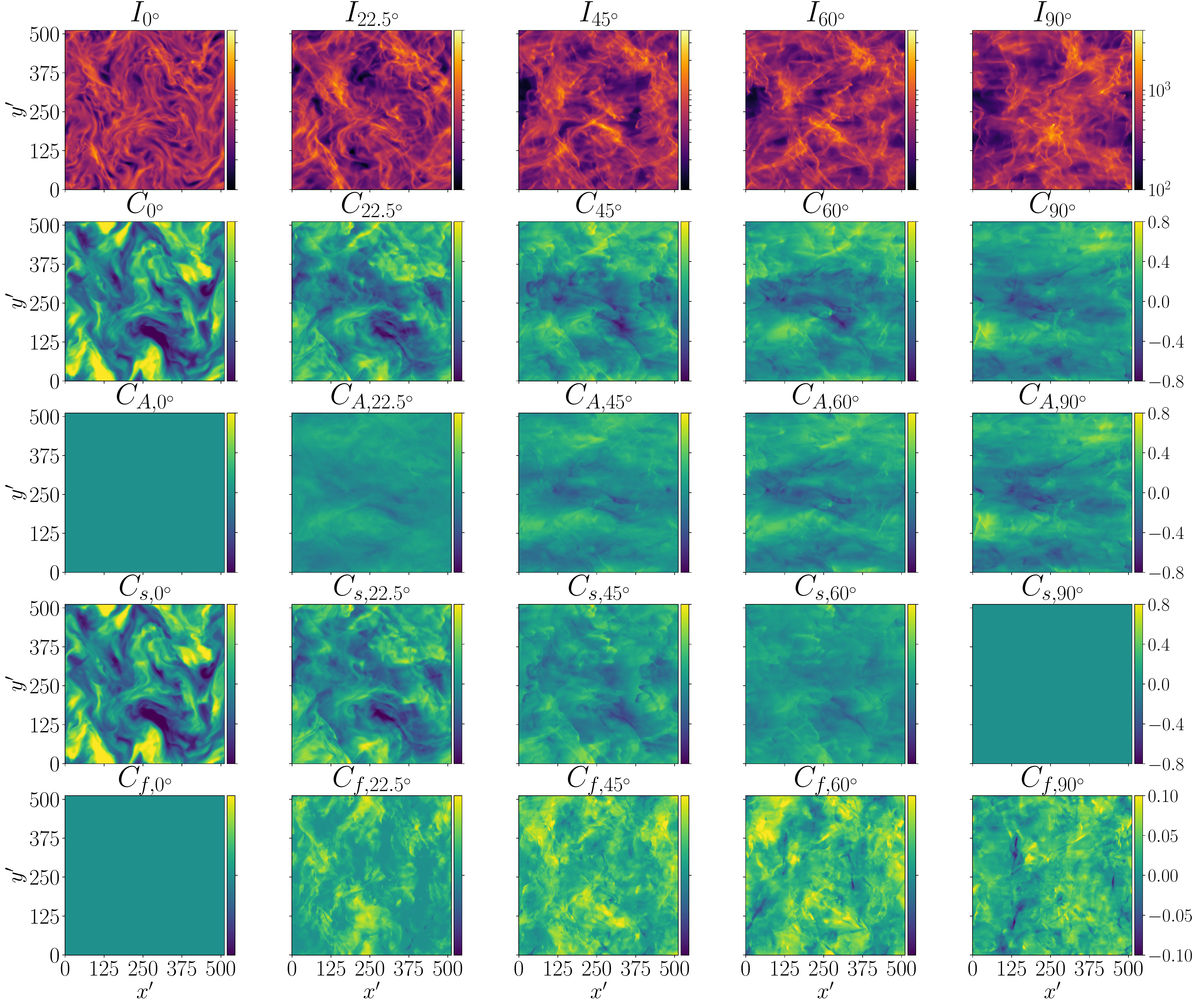}
  \caption{Two dimensional  maps obtained from one of the simulations (model M13, $\mathcal{M}_\mathrm{A}\sim 0.4$ and $\mathcal{M}_\mathrm{s}\sim 7.6$).
  By columns, from left to right, the maps obtained with $\gamma=0\degr,~22.5\degr~40\degr,~60\degr,~90\degr$ (respectively).
  By rows, from top to bottom: integrated intensity (proportional to column density, first row), centroids obtained with the original velocity (second row), centroids with the Alfv\'en mode velocity (third row), with the slow mode (fourth row), and with the fast mode velocity (fifth row). Notice that all centroids maps, except those for the fast-MHD modes, are shown with the same range of values (as indicated in the colorbar at the right. The fast modes are considerably smaller in magnitude, and thus to be visible are plotted with a smaller dynamical range.}
  \label{fig:2Dmaps}
\end{figure*}

In Figure \ref{fig:2Dmaps} we show some examples of two-dimensional maps obtained from one of the simulations (model M13 $\mathcal{M}_\mathrm{A}\sim 0.4$ and $\mathcal{M}_\mathrm{s}\sim 7.6$). Ordered by rows, from top to bottom we display the integrated intensity (column density),  centroids with the original velocity field, and centroids obtained with the Alfv\'en, slow and fast MHD velocities, respectively.
By columns we vary the viewing angle $\gamma$, from $0\degr$ (LOS $\parallel$ to $\bm{B}_0$) to $90\degr$ (LOS $\perp$ to $\bm{B}_0$).

We can see from the figure that the structures integrated parallel to the magnetic field (leftmost column) are quite isotropic. In the subsequent columns the structures become increasingly anisotropic and align with the direction of the mean magnetic field projected onto the plane of the sky (horizontally).
From the maps in the figure we see for this model (but it holds true for all) that the Alfv\'en modes dominate the contribution to the centroids when the LOS is perpendicular to the mean magnetic field. This can be readily understood by noting that Alfv\'en modes are transverse waves that travel in the direction of the magnetic field, thus the plasma displacement takes place in a perpendicular direction, thus having little plasma velocity along a LOS parallel to the mean field.  At the same time, when the LOS is parallel to the mean field the slow modes have the largest contribution.

\begin{figure*}[!htp]
	\includegraphics[width=\textwidth]{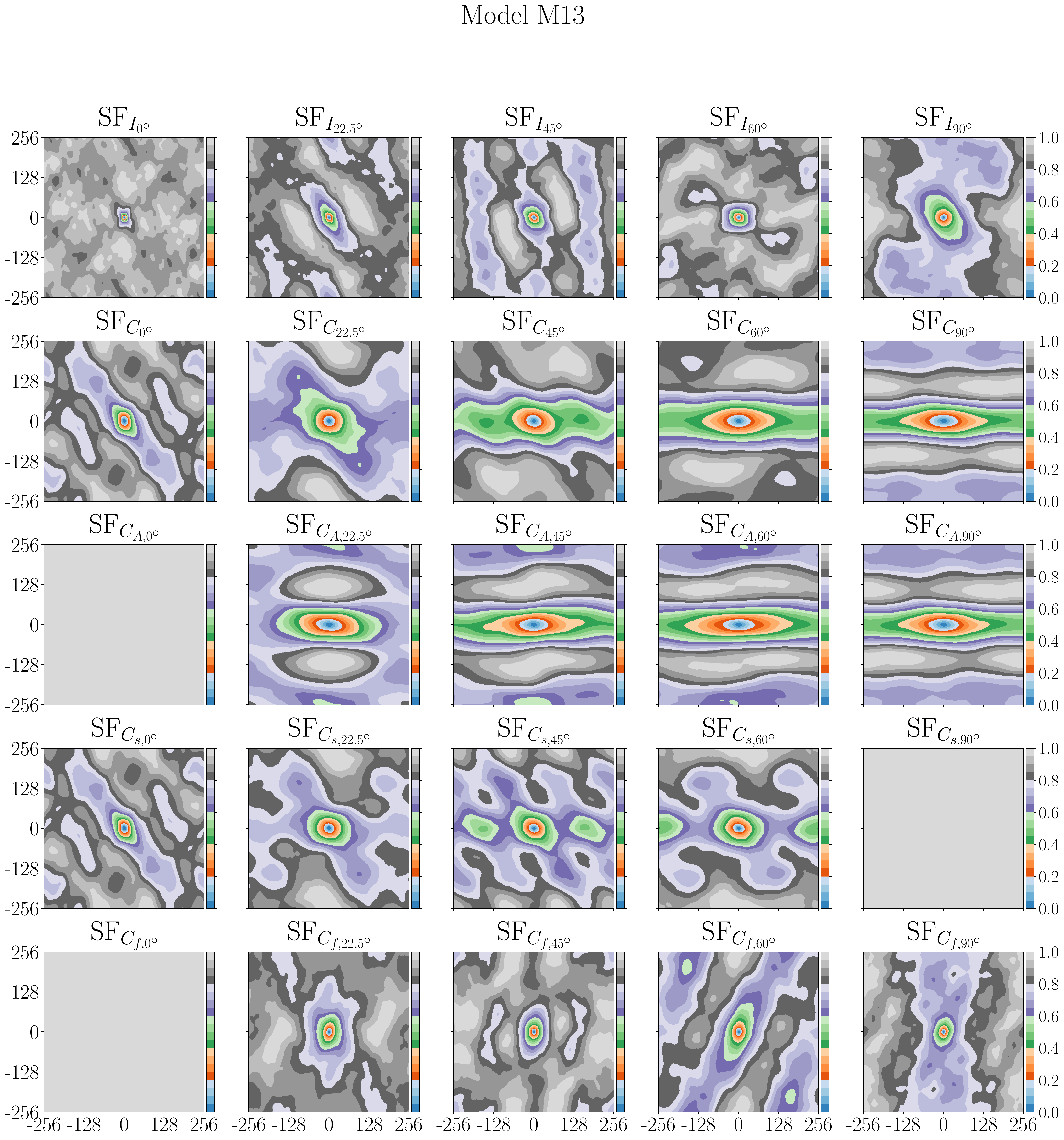}
    \caption{Two dimensional structure functions of the maps shown in Figure \ref{fig:2Dmaps} (M13, $\mathcal{M}_\mathrm{A}\sim 0.4$ and $\mathcal{M}_\mathrm{s}\sim 7.6$). The maps have the same arrangement of Figure \ref{fig:2Dmaps}, with the corresponding 2D map indicated in the title of each plot.}
    \label{fig:2dSF}
\end{figure*}

The anisotropy mentioned above is easier to characterize if we observe the structure or correlation functions. We obtained the (2D) structure function from all the two-dimensional maps, and normalize it in such a way that they lie in the range between $0$ and $1$. For instance the structure function of the column density is computed as
\begin{equation}
\label{eq:SF}
	SF_\mathrm{I_\mathrm{\gamma}}(\textbf{R}) = \frac{\langle [I_\mathrm{\gamma}\left(\bm{X}\right)-I_\mathrm{\gamma}\left(\bm{X}+\bm{R}\right)]^2\rangle}{\max \left\{\langle [I_\mathrm{\gamma}\left(\bm{X}\right)-I_\mathrm{\gamma}\left(\bm{X}+\bm{R}\right)]^2\rangle \right\}} .
\end{equation}
Where $\langle \dots \rangle$ denotes averaging over all the plane of the sky ($\bm{X}$), and the lag $\bm{R}$ is a two dimensional vector on the plane of the sky (for isotropic turbulence the structure function only depends on the magnitude of $\bm{R}$, and not on its direction).

In Figure \ref{fig:2dSF} we show as colormaps the 2D structure functions of all
the maps shown in \ref{fig:2Dmaps} (for model M13, $\mathcal{M}_\mathrm{A}\sim 0.4$ and $\mathcal{M}_\mathrm{s}\sim 7.6$). We can see from the first row in the Figure (integrated intensities) that the structure functions are somewhat anisotropic. However, the direction of alignment is neither parallel nor perpendicular to the B field.
At the same time one can see more clearly how the centroids become increasingly anisotropic as the angle between the LOS and the mean magnetic field changes from $0\degr$ to $90\degr$, aligning in the direction of the mean magnetic field (horizontally).  The model chosen for the figure has a large magnetization, so the anisotropy is quite evident, but the same trend is observed in general. We can see that at the extreme values of $\gamma$ the structure function of the original centroids is almost identical to that of the Alv\'en-mode centroids (for $\gamma = 90\degr$), or to the slow-mode centroids (for $\gamma =0\degr$). At intermediate angles the structure function of the original centroids is a combination of the two (with a marginal contribution of the fast modes), resembling more to the slow-mode centroids for low values of $\gamma$, and to the Alfv\'en modes for larger values of $\gamma$.

In order to quantify the anisotropy in the 2D structure function maps we can define (see \citetalias{2011ApJ...740..117E}) an {\it isotropy degree}:
\begin{equation}
\label{eq:ID}
\mathrm{Isotropy~degree}(\ell)= \frac{SF\left( \ell \,\bm{\hat{e}}_{\parallel}\right) }{SF\left(\ell \,\bm{\hat{e}}_{\perp} \right)},  
\end{equation}
where $\ell$ is the magnitude of the lag, which is taken in two orthogonal directions, $\bm{\hat{e}}_{\parallel}$ along the direction of the elongation of the contours (which is parallel to the magnetic field in the plane of the sky), and $\bm{\hat{e}}_{\perp}$. In the structure functions shown in Figure \ref{fig:2dSF} these two correspond roughly to horizontal for the parallel direction, and vertical for the perpendicular direction, except for the leftmost column in which both are perpendicular to the mean magnetic field.
This isotropy degree will be equal to one if the structure functions are isotropic, less than one if the contours are elongated in the direction of the magnetic field, and larger than one if their elongation occurs perpendicular to it. We must note that the anisotropy degree is different from the {\it alignment measure} used in dust grain alignment studies, and adopted in the context of the VGT \citep{2017ApJ...835...41G, 2018ApJ...865...54Y}, in which the ratio of an ellipse axes fit to a given iso-contour of correlation function was measured.

A problem, which is exclusive to numerical simulations, is the limited range of inertial range available to measure the anisotropy. The smallest scales ($\lesssim 10$ cells) are dominated by numerical diffusion and should be avoided for anisotropy studies. The largest scales should also be avoided, as they are dominated by the driving scale. The driving of the turbulence in our models is at a scale of $\sim 200$ cells. Thus we restrict our analysis to scales only between $10$ to $102$ cells (one fifth of the entire domain). However, as can be seen in Figure \ref{fig:2dSF}, even at intermediate scales the contours often become misaligned from the horizontal towards large lags. This distortion is not physical but the result of not having enough inertial range, in other words it is the effect of the turbulent cascade not having fully decoupled from the driving at such scale.
This issue was addressed in \citet{2018ApJ...865...54Y}, where a rotation of the 2D maps (correlation function in their case, structure function in ours) was performed in order to align the iso-contours horizontally (see figure 3 therein) to compensate for the distortion. We follow a similar approach, which is described in the next section.

\begin{figure*}[htp]
    \centering
	\includegraphics[scale=.57]{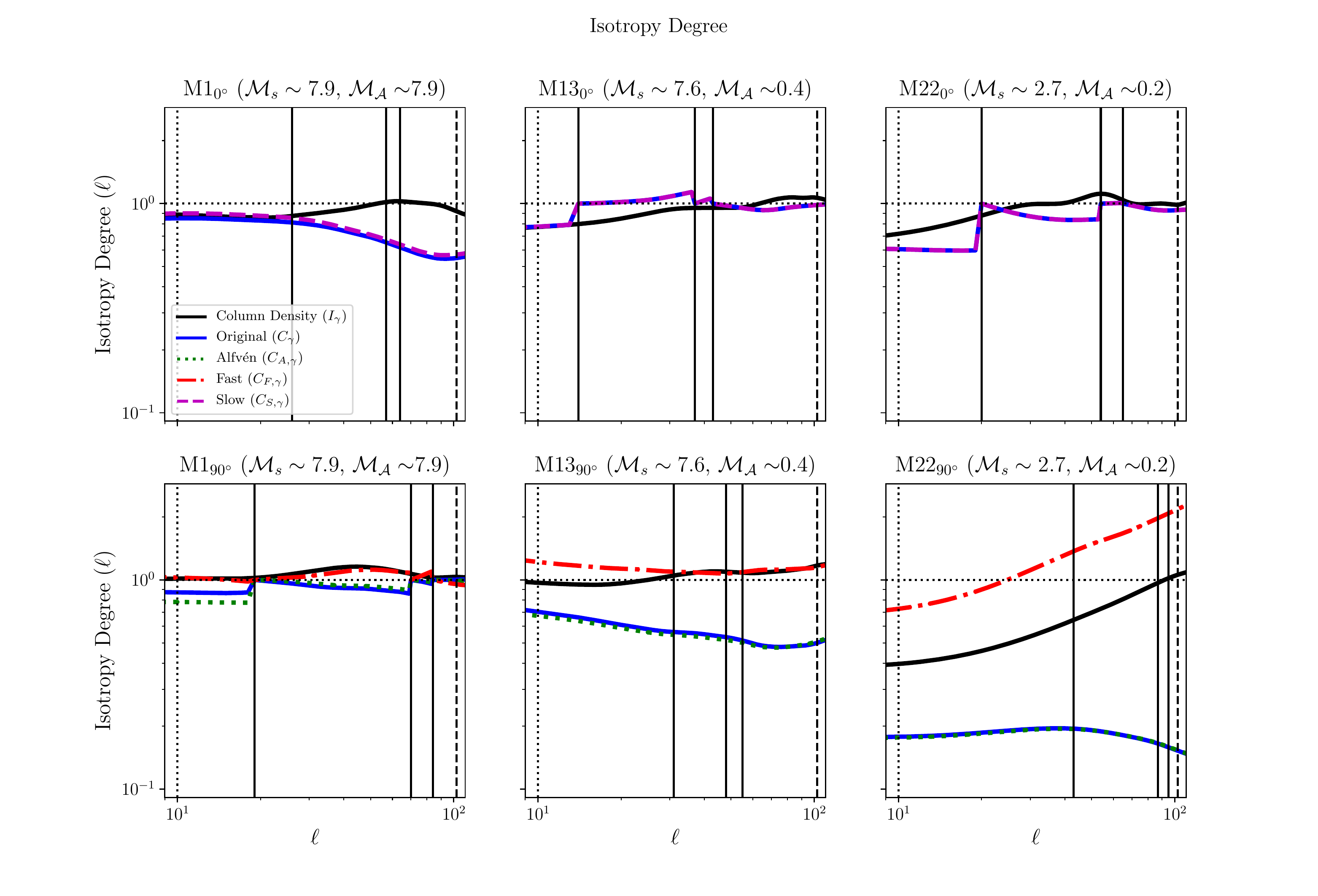}
	\caption{Isotropy degree for some selected models (indicated at the top of each plot), as a function of the scale $\ell$. The plots in the top row are obtained with PPVs with an angle of $\gamma=0\degr$ (LOS $\parallel$ to $\bm{B}_0$), and at the bottom with $\gamma=90\degr$ (LOS $\perp$ to $\bm{B}_0$).
	The different lines correspond to the centroids with the original velocity ($C_\mathrm{\gamma}$), with the Alfv\'en mode velocity ($C_\mathrm{A,\gamma}$), the fast mode ($C_\mathrm{f,\gamma}$), and the slow mode ($C_\mathrm{s,\gamma}$), as indicated in the legend in the bottom left panel. The vertical lines (dotted and dashed) denote the ranges over which the average isotropy is calculated, the dotted line at the lower end ($10$ cells), and at the upper end the dashed line (at $1/5$ of the computational box). The vertical solid lines denote the scales used to rotate the structure function to fix large scale distortions (in the iso-contours values $0.24$, $0.59$, and $0.65$), see the text for further details.
}
    \label{fig:id}
\end{figure*}

In Figure \ref{fig:id} we show examples of the isotropy degree for some of the models  with increasing magnetization from  top to bottom; on the left column observed in a LOS parallel to the mean magnetic field ($\gamma=0\degr$), and on the right column observed perpendicular to it ($\gamma=90\degr$). From the figure it is evident that the structure functions are quite isotropic if the LOS is aligned with the mean magnetic field (left column). And at the same time that the structure functions are increasingly anisotropic for higher magnetization, when the LOS is perpendicular to the average magnetic field. It is also noticeable that the isotropy degree depends only slightly on the scale ($\ell$). Also, for $\gamma=0\degr$, we see that the anisotropy in the original centroids (that is with the full --original-- velocity field, blue lines) coincides with that of the centroids obtained with the slow-mode velocity (magenta, dashed line). Whereas for $\gamma=90\degr$, the anisotropy seen in the original centroids coincides with that obtained with the Alfv\'en-mode velocity field (green, dotted line).

\section{Results}\label{sec:results}

In what follows we calculate an average isotropy degree, considering all scales that are approximately within the inertial range. As the inertial range is difficult to determine exactly from the simulations, and since it varies slightly from model to model, we consider a range between $10$ and $102$ cells (one fifth of the computational domain). However, as mentioned in the previous section, in many of the models we see a distortion of the direction of the contours towards large scales. This numerical artifact was noted by \citet{2018ApJ...865...54Y}, and we follow a similar procedure to compensate such distortion. For each model and angle $\gamma$ we take the structure function maps of velocity centroids (e.g. second row from the top in Figure \ref{fig:2dSF}) and calculate the orientation of the iso-contours at three values $0.24$, $0.59$, and $0.65$.  We then rotate all the structure function maps to align the $0.24$ iso-contour with the horizontal axis, and calculate the anisotropy degree from $\ell=10$ cells to $\ell_{0.24}$ (the scale at which the $0.24$ contour cuts the horizontal axis). We then repeat the procedure with the next two contours, obtaining the isotropy degree from $\ell_{0.24}$ to $\ell_{0.59}$, and from $\ell_{0.59}$ to $\ell_{0.65}$. We finally average the isotropy degree within these scale ranges. In the averaging procedure we restrict the largest scale to the shortest length of $102$ cells, or $\ell_{0.65}$.

\subsection{Mean velocity anisotropy}\label{sec:ResVel}

The theoretical framework put forth recently by \citetalias{2016MNRAS.461.1227K,2017MNRAS.464.3617K} is based in the structure/correlation-tensor of each of the velocity MHD modes, and an expansion in spherical harmonics of their power spectra. We must note that the bulk of their predictions consider only velocity. That is, they neglected the density structure-tensor because the expressions for the centroids become rather complicated and cumbersome. In other words, their predictions are strictly applicable to the anisotropy of mean velocity maps (see Equation \ref{eq:meanV}). However, as we pointed out earlier, these maps are can not be obtained from observations; but they can be obtained from the simulations and thus can be used to estimate the importance of including the density fluctuations.

\begin{figure*}[htp]
  \centering
	\includegraphics[width=1.0\textwidth]{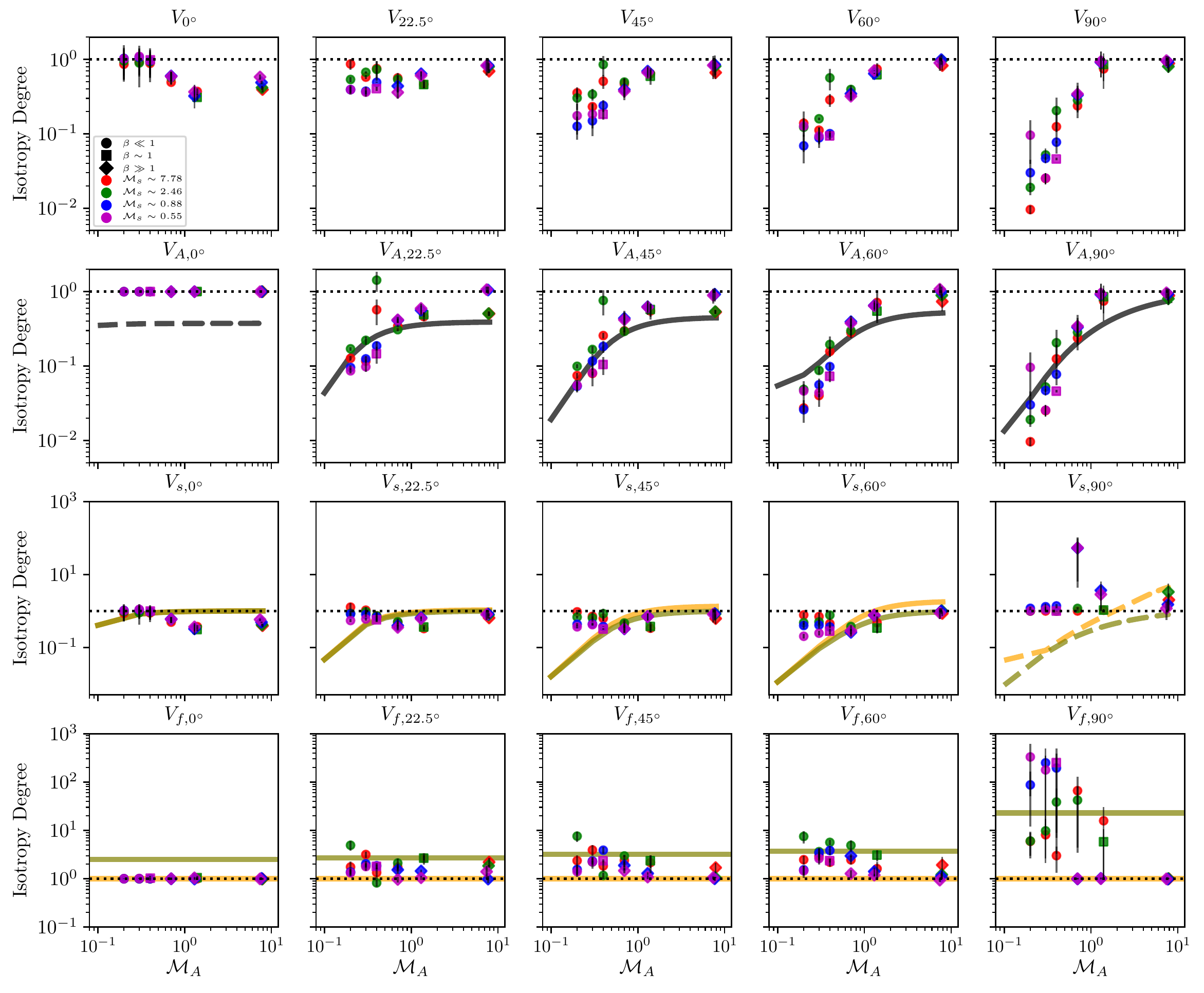}
  \caption{Average isotropy of the structure functions of mean velocity vs. the Alfv\'enic Mach number for all the models. In rows, by descending order we show the results for the original velocity field, the Alfv\'en mode, the slow mode, and the fast mode. In columns, from left to right  we vary the viewing angle, from $0$ (LOS $\parallel$ to $\bm{B}_0$), $30$, $45$, $60$, and $90\degr$ (LOS $\perp$ to $\bm{B}_0$). The shape of the symbols separate the models with high-$\beta$ (mostly hydrodynamic, diamonds), intermediate ($\beta\sim1$, squares), and low-$\beta$ (magnetically dominated, circles). The color of the symbols group models with similar $\mathcal{M}_\mathrm{s}$, as labelled in the legend at the top left panel.
  We also include the analytic predictions in \citetalias{2017MNRAS.464.3617K} for the Alfv\'en mode (black line), and the slow and fast modes (for high $\beta$ in orange and low $\beta$ in green). The anisotropy predictions for the Alfv\'en mode at $\gamma=0\degr$, and for the slow mode at $\gamma=90\degr$ are shown in dashed lines as they are only a formal limit at zero intensity of the signal (see text).}
  \label{fig:idpV}
\end{figure*}

In Figure \ref{fig:idpV} we show the average isotropy degree for maps of mean LOS velocity for all the models at five different viewing angles $\gamma$, as a function of the Alfv\'enic Mach number.
Each symbol shown in the figure shows the average isotropy degree from the structure functions calculated as described at the end of Section \ref{sec:centroids}. Each column in the figure corresponds to a different angle between the LOS and the mean magnetic field (indicated in the title of each plot). Each row corresponds to the results from different 2D maps, from top to bottom: mean (original) velocity, mean velocity obtained with the Alfv\'en mode, mean velocity of the slow-mode, and mean velocity of the fast mode. In addition we group models with similar $\mathcal{M}_\mathrm{s}$ (same $P_\mathrm{gas,0}$, see Table \ref{tab:models}) by color, and denote the $\beta$ regime with symbols of different shapes, as shown in the legend inside the top-left plot. We also include in the figure, the analytical predictions for the anisotropy from \citetalias{2017MNRAS.464.3617K}, for the Alfv\'en mode (thick black line), and the slow and fast mode (for high $\beta$ in orange, and low $\beta$ in green).
The error bars included in the figure correspond to the variability with scale (taken from the minimum and maximum of the isotropy degree within the range of scales used for averaging).

The first thing to notice from the mean LOS velocity anisotropy is a gradual increase with the viewing angle (see top row in Fig. \ref{fig:idpV}), which is more pronounced for higher magnetized runs (smaller $\mathcal{M}_\mathrm{A}$), this is consistent with previous findings in \citetalias{2011ApJ...740..117E} and \citetalias{2014ApJ...790..130B}.
By adding the velocity fields resulting from the decomposition procedure (the three MHD modes) the original velocity field is recovered, thus it is natural that the resulting anisotropy would be a combination of the anisotropy in each of the MHD modes. This can be confirmed noting that top row in Figure \ref{fig:idpV} is a combination of the bottom three rows.
However, it is important to note that different modes have different anisotropy, and different dependence with the viewing angle.
For instance, we can see that for small viewing angles ($\gamma \lesssim 30\degr$) the anisotropy of the mean velocity is dominated by that of the slow-modes. At the same time, for large viewing angles ($\gamma \gtrsim 60\degr$) the anisotropy of the Alfv\'en-mode mean velocity dominates the anisotropy of the mean (original) velocity.

We can also see that the measured anisotropies in mean LOS velocity maps are in fair agreement with the analytical predictions in \citetalias{2017MNRAS.464.3617K}.

The Alfv\'en mode anisotropies follow the trends with $\mathcal{M}_\mathrm{A}$ of the theory predictions. Note that formal theoretical anisotropy limit is non-zero
at $\gamma \to 0$, but such limit occurs at vanishing intensity of the signal and
cannot be measured in any realistic simulations or data where the isotropic
numerical or experimental noise contribution will always dominate.
A similar situation occurs for the slow mode in the limit of  $\gamma \to 90\degr$. At such viewing angle the structure function of slow modes both for high and low $\beta$ vanishes (\citetalias{2017MNRAS.464.3617K}); but interestingly the ratio of the moments is finite (e.g. they approach zero at the same rate), yielding a finite anisotropy. Such anisotropy cannot be expected to be measured realistically.

We see that the slow mode is less anisotropic than the Alfv\'en mode, and the anisotropy in our models reflects that fact, with some systematic departure at higher magnetization, where the measured anisotropy is smaller than the prediction. We must note also that the panel for $\gamma =  90\degr$ is very noisy for the fast and slow modes, as most of the velocity from the decomposition method is assigned to the Alfv\'en mode in that case. The same is true for the Alfv\'en mode at $\gamma = 0\degr$ where most of the velocity is contained in the slow MHD modes.

The fast modes have an isotropic energy spectrum, thus their anisotropy could only be due to the structure tensor. In the case of high $\beta$ the tensor structure is also isotropic, therefore the prediction is an isotropy degree of $1$. For low $\beta$ the anisotropy does not depend on the magnetization but only on the viewing angle. The correspondence between the predictions and the isotropy measured is not perfect, but we do see an increase of isotropy degree (anti-aligned with the mean magnetic field, i.\,e. $>1$) with an increase with $\gamma$ for models with $\beta < 1$.
The small amplitude of the fast modes (at all angles) makes their signal difficult to pick-up since it is tipicaly contained in only a few channels centered at zero velocity. Thus, they are most susceptible to be affected by noise and/or density fluctuations.

\subsection{Observable Anisotropies}\label{sec:obs}

To make a more realistic interpretation of the anisotropies observed, we present in this subsection the results obtained from the simulations arranged into PPV arrays. These can be directly related with observations in which both the density and velocity contribute to the emission in each velocity channel. This anisotropy has been previously studied in \citetalias{2011ApJ...740..117E} and \citetalias{2014ApJ...790..130B}, but the theory developed in \citetalias{2016MNRAS.461.1227K} and \citetalias{2017MNRAS.464.3617K} allows us to get an insight of the contribution of the different MHD modes. Unfortunately, it is not possible now to distinguish the contribution of each mode directly from observations, but by constructing centroids with their associated velocity field we are able to start looking for signatures of the compressible modes in ISM turbulence, as well as to gauge how much the density fluctuations affect the analytical predictions.

\begin{figure*}[htp]
  \centering
	\includegraphics[width=1.\textwidth]{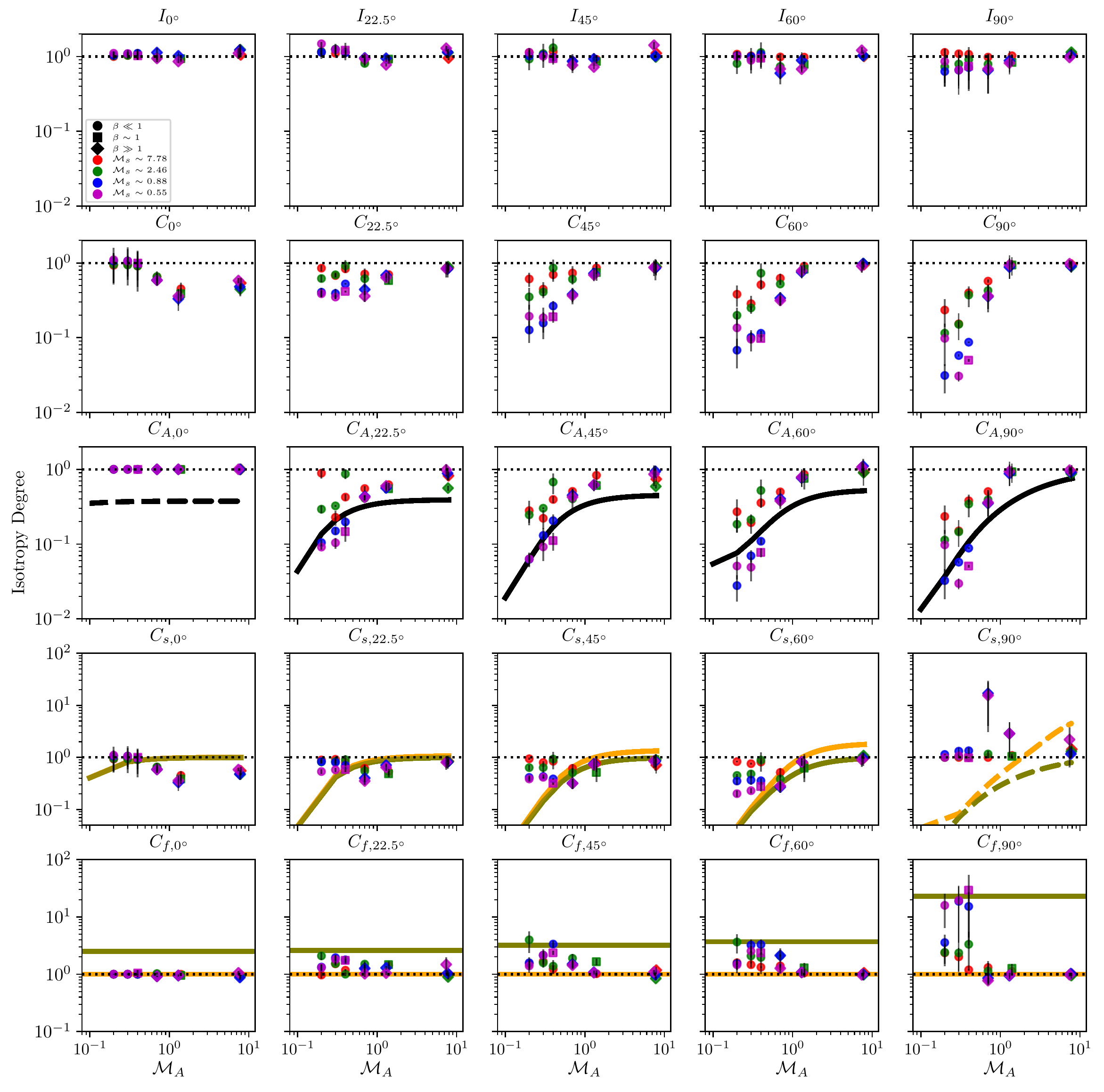}
  \caption{Same as Figure \ref{fig:idpV}, but for, in descending order by rows:
  integrated intensity, velocity centroids obtained with the original velocity, with the Alfv\'en mode velocity, with the slow mode, and with the fast mode. The analytical predictions (thick black, orange and green lines) are the same as presented in  Figure \ref{fig:idpV}.}
  \label{fig:idp}
\end{figure*}

In Figure \ref{fig:idp} we present the results in a similar arrangement as in the previous subsection (see Figure \ref{fig:idpV}), but we include an additional row at the top with the results of the integrated intensity (e.g. column density).

\subsubsection{Anisotropy in column density maps}\label{sec:ResCol}

A simple inspection shows that column density (integrated intensity) maps are qualitatively different to maps of velocity centroids, with an anisotropy that is not evident at first glance, and a richer small scale structure compared with velocity centroid maps (see e.g. Fig \ref{fig:2Dmaps}). Such small scale structures are well known to translate into a shallower power-spectra with increasing sonic Mach number \citep{2005ApJ...624L..93B,2007ApJ...658..423K}.

As for the anisotropy of the structure function maps (see Figure \ref{fig:2dSF}), they are also different from that seen in velocity centroids. For small viewing angles ($\gamma \lesssim45\degr$) column density structure functions are basically isotropic within the inertial range. For larger viewing angles the structure functions are anisotropic, with contours of equal structure aligned with the mean magnetic field. However, the degree of isotropy  is mostly independent on $\mathcal{M}_\mathrm{A}$, and only slightly dependent on $\mathcal{M}_\mathrm{s}$.
We also see some evidence of scale dependence (larger error bars) in column density maps when observed at large viewing angles.

\subsubsection{Anisotropy in velocity centroids}\label{sec:ResCent}

From a visual inspection of the velocity centroid maps in Figure \ref{fig:2Dmaps}, and from the results of with the mean velocity maps, one should expect a gradual change from a mostly isotropic structure at an observing angle $\gamma = 0\degr$, to a more anisotropic map when $\gamma=90\degr$.
Naturally, as seen previously with the mean LOS velocity, for the case of velocity centroids the map obtained with the original velocity (second row from top to bottom) is a combination of the three MHD modes (three bottom rows). And, of the three MHD modes we can see that the fast modes have the smallest amplitude (thus the smallest imprint on the observed centroids), while Alfv\'en modes dominate for large values of $\gamma$, and the slow modes for small values of $\gamma$.

This can be confirmed in the average anisotropy shown in Figure \ref{fig:idp} where the anisotropy of the centroids obtained with the slow-mode velocity traces the original centroids (with the full velocity field) for small values of the viewing angle ($\gamma \lesssim 30\degr$). At the same time, for large viewing angles ($\gamma\gtrsim 60\degr$) the anisotropy of the Alfv\'en-mode centroids dominate the anisotropy of the original centroids.

In agreement with \citetalias{2011ApJ...740..117E} and \citetalias{2014ApJ...790..130B}, the anisotropy of the velocity centroids increases with the level of magnetization (decreasing $\mathcal{M}_\mathrm{A}$), and increases also with $\gamma$,  having a maximum level of anisotropy when the LOS is perpendicular to the mean magnetic field ($\gamma=90\degr$).

The isotropy degree calculated for the velocity centroids in Figure \ref{fig:idp} is remarkably similar to that of the mean velocity maps (Figure \ref{fig:idpV}). We see a clear increase of the anisotropy with the observing angle $\gamma$. Also seen in the mean velocity maps, the isotropy in centroids at small viewing angles ($\gamma \lesssim 30\degr$) resembles that of the slow-modes, while for larger viewing angles ($\gamma \gtrsim 60\degr$) it resebles that of the Alfv\'en modes.

With the exception o the Alfv\'en mode at $\gamma=0\degr$ and the slow modes at $\gamma = 90\degr$ the isotropy is in fair agreement with the analytical predictions (\citetalias{2017MNRAS.464.3617K}. The Alfv\'en anisotropy increases with viewing angle and decreases with $\mathcal{M}_\mathrm{A}$). A similar behavior, but with a smaller anisotropy and a systematic departure at higher magnetization (lower $\mathcal{M}_\mathrm{A}$) is seen in the slow-modes. We must note that for such high magnetization the energy is dominated by the Alfv\'en mode (see Table \ref{tab:models}), and the rest of the MHD modes are confined to a small number of velocity channels and thus they become more prone to be affected by statistical noise.

It is also worth noting that, also in agreement with our previous findings, that the anisotropy in the velocity centroids is mostly \emph{scale independent} (their error bars in Figure \ref{fig:idp} are small).

\subsection{Impact of density fluctuations in velocity centroids}
\label{sec:ResMeanv}

As mentioned above the results obtained in terms of velocity centroids show the same general trends seen already with the mean LOS velocities.
 One thing to notice comparing the results from Figures \ref{fig:idpV} and \ref{fig:idp} is that the inclusion of the density fluctuations makes for a larger spread of the average anisotropy degree. In fact such spread tends to separate the data with respect to the sonic Mach number, yielding a slightly more pronounced anisotropy in subsonic models and less pronounced as the sonic Mach number increases. The reason for this is the formation of shocks in supersonic turbulence.

Since we are mostly interested in the dependence of the anisotropy in the different modes as a function of the magnetization (Alfv\'en Mach number), the plots in Figures \ref{fig:idpV} and \ref{fig:idp} are in terms of $\mathcal{M}_\mathrm{A}$ for selected angles. In addition, in appendix \ref{sec:app} we present the same results as a function of the viewing angle $\gamma$ (Figures \ref{fig:idpA-RC} and \ref{fig:idpA}), splitting the plots in columns with similar magnetization ($\mathcal{M}_\mathrm{A}$) and separating $\mathcal{M}_\mathrm{s}$ by colors. Such version of the plots allow to identify more clearly the dependence on the sonic Mach number.

\section{Summary}
\label{sec:sum}

We study the anisotropy of the structure function of the mean LOS velocity and velocity centroids in synthetic observations obtained with a grid of MHD simulations of fully developed isothermal turbulence following the methodology presented in \citetalias{2011ApJ...740..117E} and \citetalias{2014ApJ...790..130B}.
Recently, the analytical description of fluctuations of emission in PPV channels started in \citet{2000ApJ...537..720L} has been extended to study anisotropy in velocity channels (\citetalias{2016MNRAS.461.1227K}), and to velocity centroids (\citetalias{2017MNRAS.464.3617K}). In there, predictions of the anisotropy arising from different MHD modes, namely the Alfv\'en, slow, and fast modes are provided. Understanding the relative contribution of the different modes in interstellar turbulence is of particular importance, as the role of compressibility in ISM turbulence is often overlooked. In this paper we extend our previous works, with an updated grid of simulations, and study the contribution to the observed anisotropy arising from all the different MHD modes.

We use the procedure in \citet{2002PhRvL..88x5001C,2003MNRAS.345..325C} to decompose the velocity field from the simulations and obtain that corresponding to each of the three different MHD modes (Alfv\'en, slow and fast modes). With the resulting velocity fields we obtain 2D maps of centroids to analyze their structure function anisotropy.

We found, in agreement with previous results, that the structure function of velocity centroids is anisotropic, and aligns with the direction of the mean magnetic field in the plane of the sky. Such anisotropy increases with the magnitude of the projection of the field in the plane of the sky. Thus, models with higher magnetization and a small viewing angle (angle between the mean field and the line of sight) are not distinguishable from models with a lower magnetization and a larger viewing angle.

After decomposing the velocity into the three MHD modes the kinetic energy associated with the Alfv\'en and slow modes dominate the total kinetic energy, with only a minor contribution (of $\lesssim 15$ percent) contained in the fast mode (with the exception of one model that has as much as $32$ percent).

The analytical predictions in \citetalias{2016MNRAS.461.1227K,2017MNRAS.464.3617K} were obtained by considering the expected structure tensor and power spectra of the different velocity modes, but assuming that the density fluctuations are small.
Thus, as a first step we study the anisotropy in mean LOS velocity maps, which is directly comparable to the theoretical predictions, but which can not be obtained from observations.
The results are in reasonably good agreement with the the analytical predictions made in \citetalias{2017MNRAS.464.3617K}, recovering most of the general trends and the level of anisotropy for the various modes. The velocity anisotropy for large viewing angles ($\gamma\gtrsim45\degr$) is found to be determined by the Alfv\'en mode, while for smaller angles the slow-mode dominates the contribution to the velocity anisotropy.

In order to estimate how important the density fluctuations are to the centroids anisotropy, something that is now missing in the analytical description, we also study maps of integrated intensity (e.g. column density), and maps of velocity centroids. The integrated intensity maps show some anisotropy, but it is significantly smaller than that observed in velocity. Such density anisotropy also increases with the viewing angle, but is does not have an appreciable dependence on the Alfv\'en Mach number, and only a slight dependence on the sonic Mach number(see Appendix \ref{sec:app}).

Our results for the velocity centroids are very similar to those obtained with the mean LOS velocity, confirming that the density fluctuations do not change qualitatively the results. The anisotropy in the centroids, however, has a slight dependence with the sonic Mach number, which decrease (yet does not disappear) in the mean velocity maps.

\acknowledgements

DHP and AE acknowledge support from CONACYT (Mexico) grant 167611, and DGAPA-PAPIIT (UNAM) grant IG-RG 100516. AE also acknowledges support through the DGAPA-PASPA program. A.L. acknowledges the support the NSF AST 1816234 and NASA TCAN 144AAG1967. Flatiron Institute is supported by the Simons Foundation.

\appendix
\section{Anisotropy as a function of the viewing angle}
\label{sec:app}

\begin{figure*}[htp]
	\includegraphics[width=1.0\textwidth]{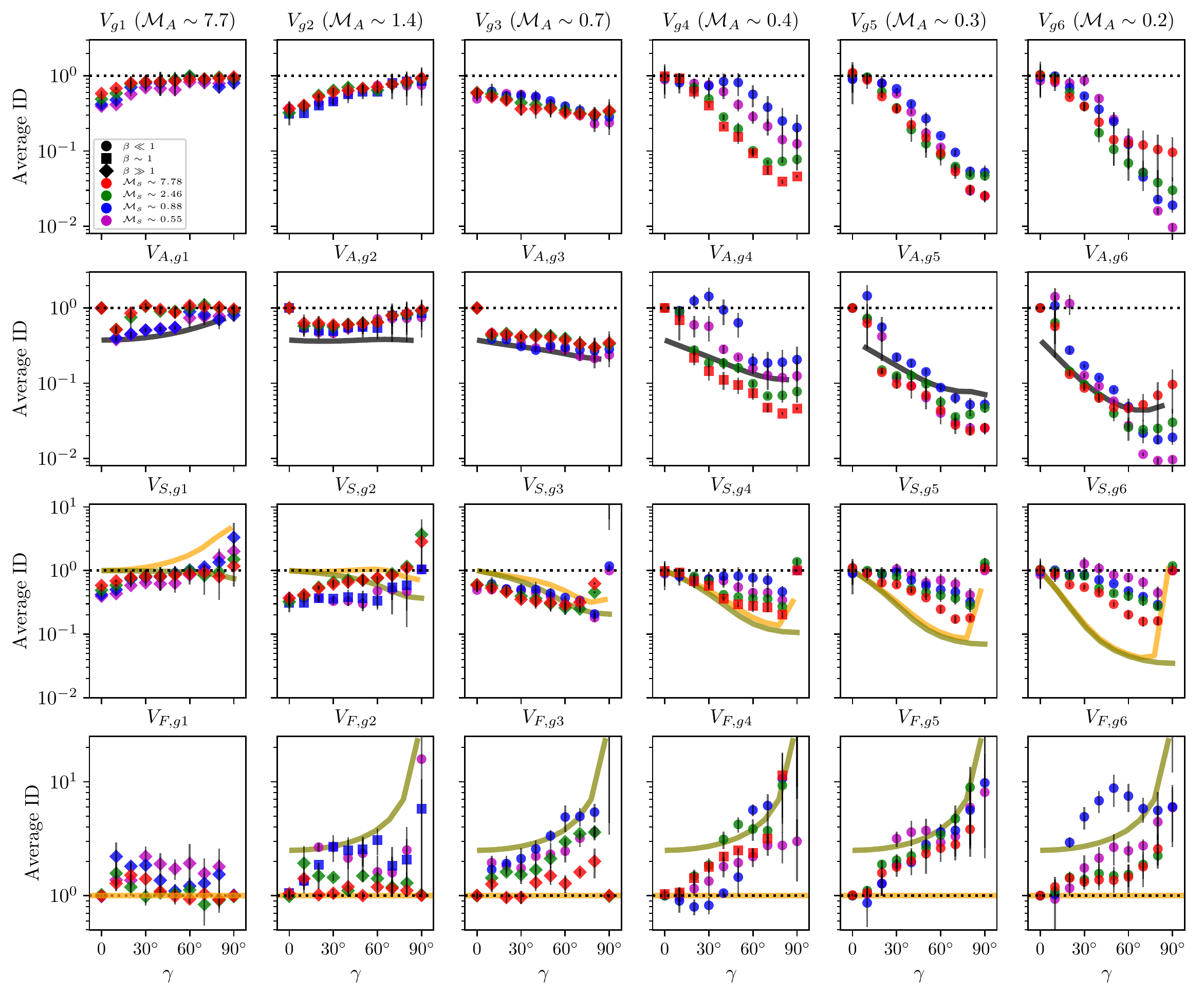}
  \caption{Average anisotropy degree of the structure functions of mean velocity (Eq. \ref{eq:meanV}) maps as a function of the viewing angle $\gamma$. In rows, from top to bottom: mean velocity maps with the original velocity field, mean velocity associated to the Alfv\'en mode, mean velocity of the slow mode, and mean velocity of the fast mode. We group in columns the models based in their magnetization, from highly super-Alfv\'enic in the left, to sub-Alfv\'enic in the right ($\mathcal{M}_\mathrm{A}$ is indicated in the title of each column).}
  \label{fig:idpA-RC}
\end{figure*}

\begin{figure*}[htp]
	\includegraphics[width=1.0\textwidth]{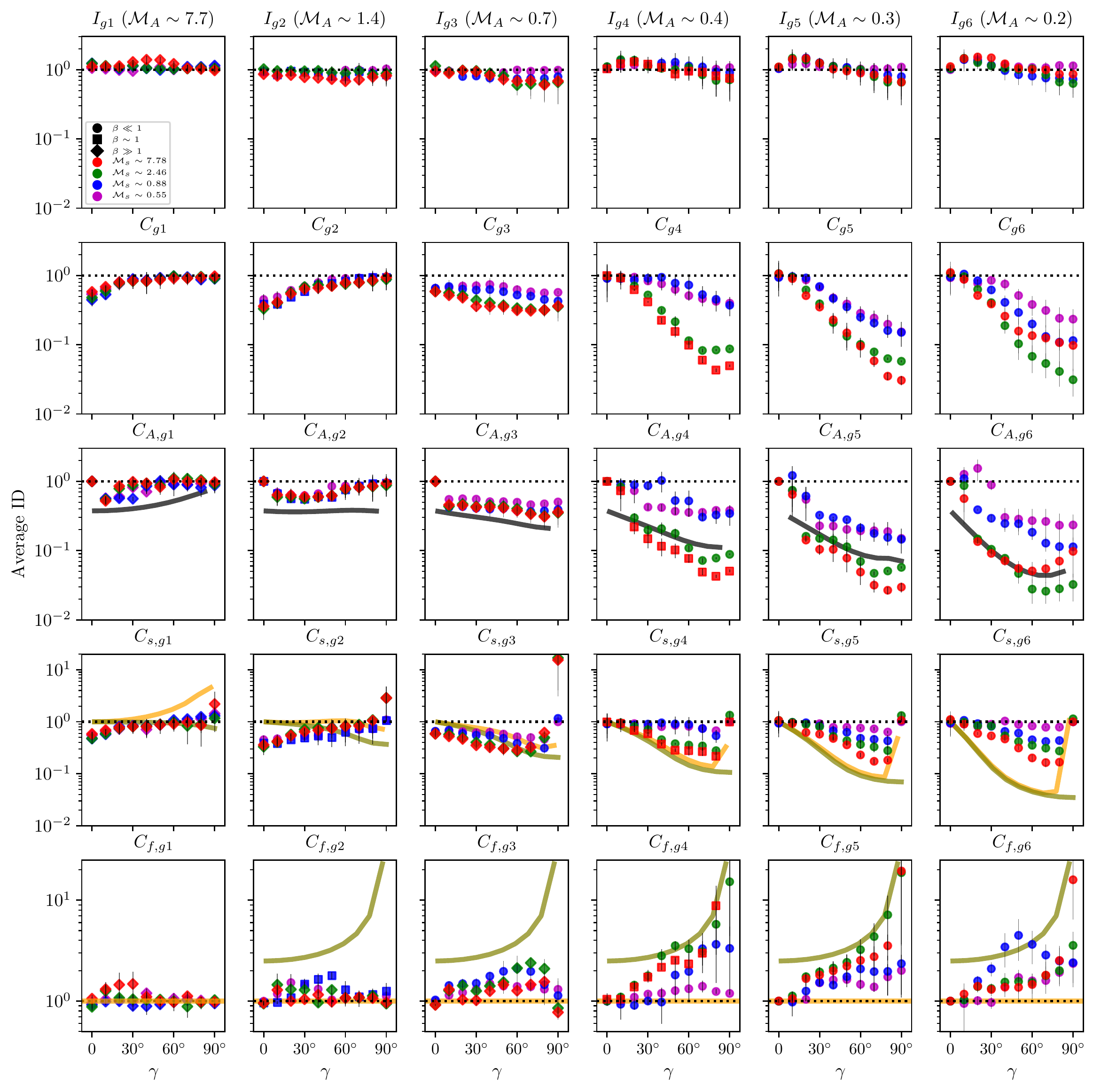}
	\caption{Average isotropy degree as a function of the viewing angle $\gamma$. The top row shows the results for the integrated intensity. In the subsequent rows (in downward progression) we show the results for the different velocity centroids, starting with the original velocity field, and then the Alfv\'en, slow, and fast modes, respectively. The arrangement in columns is the same as in Figure \ref{fig:idpA-RC}. We also include the analytic predictions in \citetalias{2017MNRAS.464.3617K} for the Alfv\'en mode (black line), and the slow and fast modes (for high $\beta$ in orange and low $\beta$ in green).}
  \label{fig:idpA}
\end{figure*}

In Section \ref{sec:results} we presented the isotropy degree in all the models in groups (columns) of viewing angle, and as a function of $\mathcal{M}_\mathrm{A}$ to highlight the dependence on the Magnetization.
In this appendix we present the same results grouping in columns by similar
$\mathcal{M}_\mathrm{A}$, as a function of the viewing angle, and the symbols colored by $\mathcal{M}_\mathrm{s}$. The anisotropy degree for the mean LOS velocity can be found in Figure \ref{fig:idpA-RC}, and for the integrated intensity and the various velocity centroids in Figure \ref{fig:idpA}
Similarly to the results presented in the main body of the paper we include the \citetalias{2017MNRAS.464.3617K} predictions.

From Figure \ref{fig:idpA-RC} we can see how the models trace the general shape of the analytical predictions. We must note however, that implicitly, the MHD mode decomposition and particular form of anisotropy relations dependent on   $\mathcal{M}_\mathrm{A}$ was assuming a value than or in the range of $1$.

In the synthetic observations the lowest magnetization models are essentially isotropic. However, the theory prediction does not have perfect anisoptropy for the Alfv\'en mode. The reason is that at high $\mathcal{M}_\mathrm{A}$ the energy spectrum is isotropized, but the structure tensor remains anisotropic.
The high beta slow modes show isotropy degree less than unity for small $\mathcal{M}_\mathrm{A}$, and larger than unity for large $\mathcal{M}_\mathrm{A}$. This can be attributed to the tensor structure and energy spectrum being of different nature (\citetalias{2016MNRAS.461.1227K,2017MNRAS.464.3617K}). The quadrupole moment due to tensor structure alone is positive (assuming isotropic energy spectrum), whereas the quadrupole moment due to energy spectrum alone (assuming isotropic tensor) is negative. At low $\mathcal{M}_\mathrm{A}$, the energy spectrum dominates the nature, whereas at high  $\mathcal{M}_\mathrm{A}$ tensor structure dominates the nature of isotropy degree . Fast modes have isotropic energy spectrum, so anisotropy is due to tensor structure only. The tensor structure for high $\beta$ fast mode is isotropic, so isotropy degree is always $\sim 1$. For low $\beta$ fast modes, the isotropy degree only depends on the viewing angle.

The anisotropy seen in velocity centroids (Figure \ref{fig:idpA}) is very similar to that of the mean LOS velocity (Figure \ref{fig:idpA-RC}). The noticeable differences are a small, but more clear separation with the sonic Mach number.

\bibliography{Master}



\end{document}